\documentclass[aps,prb,superscriptaddress,reprint]{revtex4-1}%
\usepackage{epsfig,pslatex,latexsym,times,amssymb,amsmath,graphicx}
\usepackage{bm, float, lipsum, bbold}
\usepackage{amsmath}
\usepackage{amsfonts}
\usepackage{amssymb}
\usepackage{mathrsfs}
\usepackage{color}
\usepackage{hyperref}
\usepackage{graphicx}%
\usepackage{gensymb}

\providecommand{\U}[1]{\protect\rule{.1in}{.1in}}
\begin{document}
\author{N. J. Harmon}
\email{nicholas-harmon@uiowa.edu}
\affiliation{Department of Physics and Astronomy and Optical Science and Technology Center, University of Iowa, Iowa City, Iowa
52242, USA}
\author{T. A. Peterson}
\affiliation{School of Physics and Astronomy, University of Minnesota, Minneapolis, Minnesota
55455, USA}
\author{C. C. Geppert}
\affiliation{School of Physics and Astronomy, University of Minnesota, Minneapolis, Minnesota
55455, USA}
\author{S. J. Patel}
\affiliation{Department of Materials, University of California, Santa Barbara, California 93106, USA}
\author{C. J. Palmstr{\o}m}
\affiliation{Department of Materials, University of California, Santa Barbara, California 93106, USA}
\affiliation{Department of Electrical and Computer Engineering, University of California, Santa Barbara, California 93106, USA}
\author{P. A. Crowell}
\affiliation{School of Physics and Astronomy, University of Minnesota, Minneapolis, Minnesota
55455, USA}
\author{M. E. Flatt\'e}
\affiliation{Department of Physics and Astronomy and Optical Science and Technology Center, University of Iowa, Iowa City, Iowa
52242, USA}
\date{\today}
\begin{abstract}
The hyperfine field from dynamically polarized nuclei in $n$-GaAs is very spatially inhomogeneous, as the nuclear polarization process is most efficient near the randomly-distributed donors. Electrons with polarized spins traversing the bulk semiconductor will experience this inhomogeneous hyperfine field as an effective fluctuating spin precession rate, and thus the spin polarization of an electron ensemble will relax.  A theory of spin relaxation based on the theory of random walks is applied to such an ensemble precessing in an oblique magnetic field, and the precise form of the (unequal) longitudinal and transverse spin relaxation analytically derived. To investigate this mechanism, electrical three-terminal Hanle measurements were performed on epitaxially grown Co$_2$MnSi/$n$-GaAs heterostructures fabricated into electrical spin injection devices. The proposed anisotropic spin relaxation mechanism is required to satisfactorily describe the Hanle lineshapes when the applied field is oriented at large oblique angles.
\end{abstract}
\title{Anisotropic spin relaxation in \emph{n}-GaAs from strong inhomogeneous hyperfine fields produced by the dynamical polarization of nuclei}
\maketitle

\begin{figure}[ptbh]
    \begin{centering}
        \includegraphics[scale = 0.38,trim =130 240 16 20, angle = -0,clip]{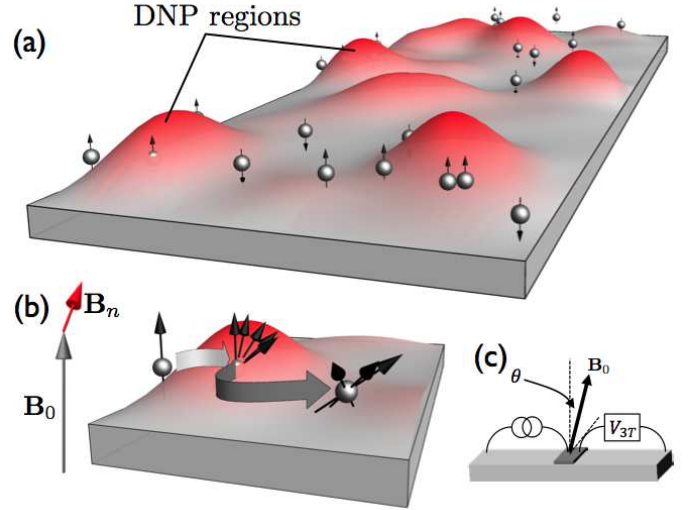}
        \caption[]{(Color online) (a) electron spins (spheres with spin vectors) in $n$-doped GaAs.  Bumpy (red) regions depict the presence of donors and nuclear fields generated by dynamic nuclear polarization. (b) The spin rotation caused by one electron spin entering and departing a DNP region. (c) Experimental geometry with $\theta$ being the angle between the applied field and sample normal.}\label{fig:diagram}
    \end{centering}
\end{figure}

\emph{Introduction}. --- The understanding of  electrical injection and detection of spin in ferromagnetic/semiconductor devices has progressed significantly over the past decade.\cite{Lou2006, Crowell2012}
A key obstacle for interpreting spin transport experiments near the metal-insulator transition has been the complicating presence of dynamically polarized nuclear spins.\cite{Affouda2009, Salis2009, Chan2009} In the process of dynamic nuclear polarization (DNP), the electron spin polarization, maintained out of equilibrium optically or electrically, is transferred to the nuclear system over long time scales via the hyperfine interaction\cite{Paget1977,
Meier1984,Kikkawa2000,Salis2001,Falk2015}, and can induce nuclear fields up to 5.3~T in GaAs.
The nature and distribution of the electronic states controls the properties of the resulting effective hyperfine fields from DNP;  for instance, electron spins in itinerant states interact rapidly with a multitude of nuclei, which dilutes the effect and leads to inefficient nuclear polarization.
Spins situated at impurity sites, however, interact with many fewer nuclei, which promotes a more efficient\cite{Paget1977} DNP.
At the doping levels examined here, the different donor wave functions overlap often but do not completely fill the bulk crystal, which consequently results in a high degree of nuclear field inhomogeneity  [see Figure \ref{fig:diagram}].\cite{Huang2012, Christie2014}
Previous descriptions of the spin transport dynamics in $n$-doped semiconductors with spin drift-diffusion equations\cite{Chan2009, Flatte2000b,Awschalom2002,Yu2002b,Crooker2005a,Crooker2005b} have neglected this essential inhomogeneity of the nuclear field.

Here we predict a new anisotropic spin relaxation mechanism in semiconductors that occurs when inhomogeneous effective magnetic fields are present, such as arise from polarized nuclei. Intermediately $n$-doped GaAs, under the conditions of DNP, offers a testbed for our theory where the inhomogeneity manifests itself as a bipartite field with values $\bm{B}_0$ or $\bm{B}_0 + \bm{B}_N$ with $\bm{B}_0$ being an applied magnetic field and  $\bm{B}_N$ the nuclear field induced by DNP near a donor. We demonstrate such a system experimentally and show that measurements of the steady state spin polarization are consistent with the devised inhomogeneity-induced anisotropic spin relaxation mechanism.

\emph{Theory}. --- The theoretical description presented herein can be understood qualitatively by examining Figure \ref{fig:diagram}, wherein the inhomogeneity of the magnetic field is represented by random placement of distortions (in red) signifying both the presence of a donor atom and a DNP-induced nuclear field. Electron spins cross between these donor regions and regions in between donors, which lack the DNP-induced nuclear field. This transit acts on the spin similarly to an effective fluctuating Zeeman field. The lower part of Figure \ref{fig:diagram} shows how the inhomogeneity relaxes the spin; in general the nuclear field is non-collinear with the applied field, which causes the precession axis to stochastically modulate when the spin changes field regions. When the transit time between these regions is much faster than the change in precession rate experienced by the spin upon transit, then the regime of spin relaxation corresponds to the motional narrowing regime. We present a general form of the theory of spin relaxation in the inhomogeneous nuclear field produced by DNP, and specifically within the motional narrowing regime  we obtain compact analytic results that can be readily incorporated into  spin drift-diffusion theories.

We now present a calculation of the spin relaxation of a spin ensemble, $\bm{S}$, ensuing from the aforementioned theory and assumptions.
In other words we would like to solve for the spin relaxation due to the following precession:
\begin{equation}\label{eq:spinDiffEq}
\frac{d\bm{S}(t)}{dt} = \gamma[\bm{B}_0 + \bm{B}_n(t) ]\times \bm{S}(t),
\end{equation}
where the spatial inhomogeneity of the nuclear field is written as a time-dependent nuclear field that takes on only two possible values of either $\bm{b}_n$ or $0$.
Since $\bm{B}_n(t)$ changes rapidly, the first approximation is to replace it with its average value: $\langle \bm{B}_n (t) \rangle = \bm{b}_n/2$ with
\begin{equation}\label{eq:nuclearField}
\bm{b}_n = b_{nuc} \langle \bm{I} \rangle =  \frac{b_{nuc} \bm{S} \cdot (\bm{B}_0 + b_e \bm{S})}{|\bm{B}_0 + b_e \bm{S}|^2 + \xi B_{\ell}^2}(\bm{B}_0 + b_e \bm{S}),
\end{equation}
where $\bm{I}$ is the nuclear spin, $b_{nuc}$ is the Overhauser coefficient, $b_e$ is the Knight coefficient, and $\sqrt{\xi}B_{\ell}$ denotes the strength of the random local field. The Knight field allows the nuclear field to be non-collinear to the applied field.

Since the average nuclear field is static, that alone will not relax the spin; temporal fluctuations around the average are required:
\begin{equation}\label{eq:spinDiffEq1}
\frac{d\bm{S}(t)}{dt} = \gamma[\bm{B}_0 + \frac{1}{2}  \bm{b}_n+ \frac{1}{2}\bm{b}_n f(t) ]\times \bm{S}(t),
\end{equation}
where $f(t)$ is a stochastic function. $f(t)$ is equal to $+1$ ($-1$) for an average time interval $1/k_n$ ($1/k_0$), where $1/k_0$ ($1/k_n$) is the average time the spin experiences the field $\bm{B}_0$ ($\bm{B}_0 + \bm{B}_n$) before that field changes.
We would like to find the dissipative effects from the time-dependent field so we will ignore the static applied field and average nuclear field:
\begin{equation}\label{eq:spinDiffEq2}
\frac{d\bm{S}(t)}{dt} = \frac{\gamma}{2} f(t) \bm{b}_n \times \bm{S}(t) = f(t) \bm{\Omega} \cdot \bm{S}(t),
\end{equation}
where $\bm{\Omega}(t)$ is the skew-symmetric matrix
\begin{eqnarray}
\bm{\Omega} &=&  \frac{\gamma}{2} b_n  \hat{\bm{\Omega}} = \frac{\gamma}{2} b_n
\left( {\begin{array}{ccc}
 0 & -\omega_z  & \omega_y  \\
  \omega_z & 0  & -\omega_x \\
 -\omega_y & \omega_x  & 0  \\
 \end{array} } \right) {}\nonumber\\
 &\equiv&{} \frac{\gamma}{2} b_n
 \left( {\begin{array}{ccc}
 0 & -\cos\alpha  & \sin\alpha \sin\beta  \\
  \cos\alpha & 0  & -\sin\alpha \cos\beta \\
 -\sin\alpha \sin\beta& \sin\alpha \cos\beta  & 0  \\
 \end{array} } \right),
\end{eqnarray}
where $\alpha$ and $\beta$ are the spherical coordinates of the nuclear field.
Depending on the value of $f(t)$, the solution to the precession equation in between field switchings is
\begin{equation}\label{}
\bm{S}(t) = e^{\bm{\Omega}t} \cdot \bm{S}_0, \quad \bm{S}(t) = e^{-\bm{\Omega}t} \cdot \bm{S}_0,
\end{equation}
where $\bm{S}_0$ is the initial spin vector.

The time evolution of the spin ensemble can be computed by the theory of continuous-time-random-walks.\cite{Montroll1965, Scher1973, Kehr1978, Hayano1979, Czech1989,Harmon2013a, Harmon2014b, Yaouanc2011}
The difficulty of the theory is reduced since the field modulates between only two values.\cite{Borgs1991,Borgs1995}
The polarization function is a result of random walks between the two spin environments:
\begin{widetext}
\begin{eqnarray}\label{eq:volterra}
\bold{P}(t) &=&  \frac{1}{2} \Bigg[ e^{\bm{\Omega}t} \Phi_{n}(t)+ \int_0^{t} \Phi_{0}(t-t')e^{-\bm{\Omega}(t-t')}\Psi_{n0}(t')e^{\bm{\Omega}t'}  dt' + \int_0^{t} \int_0^{t' } \Phi_{n}(t-t')e^{\bm{\Omega}(t-t')}\Psi_{0n}(t'-t'')e^{-\bm{\Omega}(t'-t'')} \Psi_{n0}(t'')e^{\bm{\Omega}t''}  dt'' dt'{}\nonumber\\
&+&{}... + \text{signs of $\bm{\Omega}$ switched and $n \leftrightarrow 0$}
\Bigg]\cdot \bm{S}_0,\nonumber
\end{eqnarray}
\end{widetext}
where $\Psi_{ij}$ are wait-time distributions to transition between state $i$ to state $j$, and $\Phi_i$ are the survival probabilities in state $i$.
Using exponential wait-time distributions leads to:
\begin{eqnarray}\label{eq:volterra2}
\bm{P}(t) &=&  \frac{1}{2} \Bigg[ e^{(\bm{\Omega}-k_n)t} +  \int_0^{t} e^{(-\bm{\Omega}-k_0)(t-t')} k_n e^{(\bm{\Omega}-k_n)t'}  dt'{}\nonumber \\
&+&{} \int_0^{t} \int_0^{t' }e^{(\bm{\Omega}-k_n)(t-t')} k_0 e^{(-\bm{\Omega}-k_0)(t'-t'')} k_n e^{(\bm{\Omega}-k_n)t''}  dt'' dt'{}\nonumber\\
&+&{}... + \text{signs of $\bm{\Omega}$ switched  and $n \leftrightarrow 0$}
\Bigg]\cdot \bm{S}_0.
\end{eqnarray}
Utilizing the Laplace transform and its convolution properties, the polarization function in the Laplace domain simplifies to
\begin{equation}\label{}
\tilde{\bm{P}}(s) = \frac{1}{2} \Big[ ( \tilde{\bold{R}}_0 + \tilde{\bold{R}}_n   +  ( k_0 + k_n ) \tilde{\bold{R}}_0 \tilde{\bold{R}}_n )      \Big] \sum_{j = 0}^{\infty} (k_0 k_n \tilde{\bold{R}}_0 \tilde{\bold{R}}_n )^j \cdot \bm{S}_0,
\end{equation}
with
\begin{equation}\label{}
\tilde{\bold{R}}_{0(n)} = \frac{1}{s + k_{0(n)} \pm \bm{\Omega}},
\end{equation}
which has a Laplace transform equal to
\begin{equation}\label{eq:M}
\tilde{\bm{P}}(s) = \frac{s +  k_0+k_n }{s(s +  k_0 + k_n)  - \bold{A}}\cdot \bm{S}_0 = \tilde{\bold{M}}(s) \cdot \bm{S}_0.
\end{equation}
where
\begin{equation}\label{}
\bold{A} = -(k_n - k_0) \bm{\Omega} + \bm{\Omega}\bm{\Omega}.
\end{equation}
This general expression can be analytically transformed to the time domain.\cite{supp}

We now apply the approximation of fast transitions, $k_{0,n}>>\gamma B_n$. To leading order in $s$, $\tilde{\bold{M}}(s)  = [s\mathbb{1} - \bold{A}/(k_0 + k_n)]^{-1}$ which is inverted to be $\bold{M}(t)  = e^{\bold{A} t/(k_0+k_n)}$ and then $\dot{\bm{P}}(t) = \frac{\bold{A}}{k_0 + k_n} \bm{P}(t)$. The next order correction yields\cite{supp}
\begin{equation}
\dot{\bm{P}}(t) = \big(\frac{\bold{A}}{k_0 + k_n}  - \frac{\bold{AA}}{(k_0 + k_n)^3}\big) \bm{P}(t)
\end{equation}
which when written out to second order in $\bm{\Omega}$ becomes
\begin{eqnarray}\label{eq:damping}
\frac{d \bm{P}(t)}{dt}
& =&  -\frac{1}{4}\frac{\gamma^2}{k_0 + k_n}\Big[ 1 - \Big(\frac{k_n-k_0}{k_0 + k_n}\Big)^2\Big]  \bm{b}_n \times (\bm{P}(t)\times \bm{b}_n) \nonumber\\
&-&{} \frac{\gamma}{2} \frac{k_n-k_0}{k_0 + k_n} \bm{b}_n \times \bm{P}(t).
\end{eqnarray}
Only the first term has the capability to relax the spin ensemble. The second term is a correction to the Larmor precession.

By combining spin effects such as spin injection, other spin relaxation sources, and adding back in the applied and average nuclear field in Eq. (\ref{eq:spinDiffEq1}), we can write the following equation to encompass the (non-diffusive) spin evolution:
\begin{eqnarray}\label{eq:main}
\frac{d \bm{P}(t)}{dt}
& =& \gamma[\bm{B}_0 + \frac{k_0}{k_0 + k_n}  \bm{b}_n ]\times \bm{P}(t)\nonumber\\
&-&{} \frac{1}{\tau_s}\bm{P}(t)  -\gamma^2 \tau  \bm{b}_n \times (\bm{P}(t)\times \bm{b}_n) +\bm{G},
\end{eqnarray}
where
\begin{equation}\label{}
\tau
 =  \frac{1}{4}\frac{1}{k_0 + k_n}\Big[ 1 - \Big(\frac{k_n-k_0}{k_0 + k_n}\Big)^2\Big],
\end{equation}
$\bm{G}||\hat{x}$ is the spin generation vector, and $\tau_s$ is other spin relaxation mechanisms which we assume to be isotropic. We have simulated the spin evolution with a Monte Carlo approach and found agreement with solutions to the differential equation (\ref{eq:main}).\cite{supp}

\emph{Experiment}. --- To test the theory, we probed the spin polarization $\mathbf{P}$ in
\textit{n}-GaAs using electrical Hanle measurements in a standard three-terminal (3T) configuration.
The sample used was an epitaxially grown Co$_2$MnSi/\textit{n}-GaAs (100) heterostructure. A 2.5
$\mu$m thick Si-doped $n = 4\times 10^{16}$ cm$^{-3}$ \textit{n}-GaAs channel was grown on an
insulating GaAs (100) substrate. To thin the naturally occurring Schottky barrier and create a tunnel
barrier for efficient spin injection\cite{Hanbicki2003}, a 15 nm \textit{n} $\rightarrow$
\textit{n}$^+$ transition layer (\textit{n}$^{} = 5 \times 10^{18}$ cm$^{-3}$) was grown followed by a
18 nm \textit{n}$^+$ layer. 5 nm of ferromagnetic (FM) Heusler alloy Co$_2$MnSi was then grown,
followed by Al and Au capping layers.

The structures were patterned into lateral spin injection devices\cite{Lou2007} using standard
photolithographic techniques.  The injection contact was 5 $\mu$m $\times$ 50 $\mu$m. Spin was
electrically injected into the \textit{n}-GaAs channel by imposing a DC current bias (800~A/cm$^2$ at 0.51~V) across the
FM/\textit{n}-GaAs interface.  For the measurements discussed here, the interface was forward-biased, so that electrons flowed
from the semiconductor into the ferromagnet.  In a 3T measurement the FM/\textit{n}-GaAs interface
voltage is measured by measuring the voltage with respect to a remote contact outside of the charge
current path. The spin polarization in the channel directly below the injection contact was probed by
measuring the change in the 3T voltage $\Delta V_{3T}$ upon application of an external out-of-plane
magnetic field $B$. This transverse magnetic field serves to precess the spins and destroy the spin
polarization in the channel via the Hanle effect\cite{Lou2006}.

The influence of DNP on the spin polarization in our samples is most clearly seen by measuring the 3T
Hanle effect when the applied field is tilted at small oblique angles $\theta$ away from the vertical direction and toward the easy
axis of the ferromagnetic contact, as shown in Fig.~\ref{fig:diagram}(c).  The oblique geometry allows for a significant hyperpolarization of the nuclei (Overhauser effect).
Satellite peaks are then
observed that correspond to fields at which the applied dephasing transverse field is partially
cancelled by the Overhauser field\cite{Chan2009}.  (A less prominent satellite peak at very low fields is due to the Knight field of the polarized electrons.)  The effectiveness in
reproducing the oblique 3T Hanle lineshapes therefore serves as a test of the validity of the model
used to account for the affects of DNP.

\begin{figure}[ptbh]
    \begin{centering}
        \includegraphics[scale = 1,trim =0 0 0 0, angle = -0,clip]{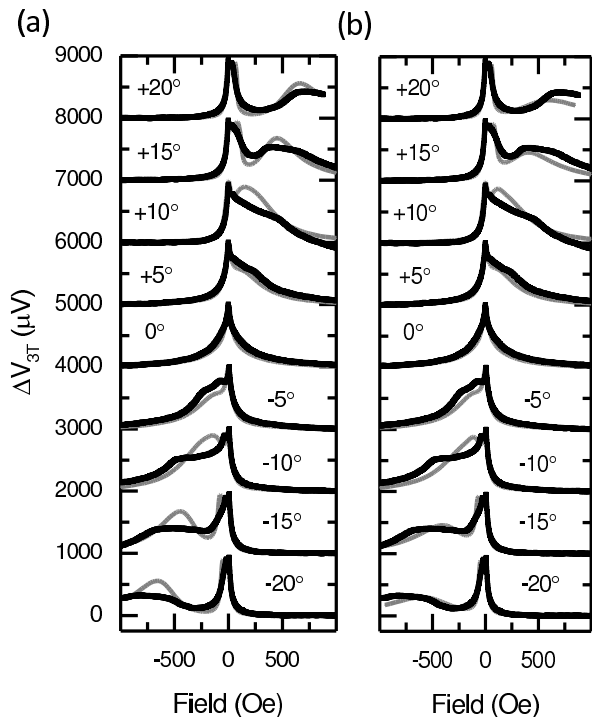}
        \caption[]{Shown is the oblique-angle 3T Hanle signal measured at 60 K. The black lines are the experimental data, with a second order (magnetoresistance) background removed and the different angles artificially offset. Shown in gray is the numerical solution to Eq. \ref{eq:spinDiffusion} for the device geometry, (a) without the anisotropic hyperfine relaxation terms and (b) with the anisotropic hyperfine relaxation terms included. One set of fitting parameters was used to simultaneously fit all the angles. These fitting parameters for both situations are shown in Table I. The anisotropic hyperfine terms improved the fit the most for larger oblique angles.}\label{fig:angleExp}
    \end{centering}
\end{figure}

\emph{Discussion}. ---
Thus far we have only examined the non-diffusive dynamics. However the importance of spin diffusion on Hanle curves has been well-documented.\cite{Lou2007}
In light of the theory hitherto presented, we write the following spin diffusion equation:
\begin{widetext}
\begin{eqnarray}\label{eq:spinDiffusion}
\frac{d \bm{P}(t)}{dt}
 = \gamma[\bm{B}_0 + \frac{k_0}{k_0 + k_n}  \bm{b}_n ]\times \bm{P}(t)
- \frac{1}{\tau_s}\bm{P}(t)  -\gamma^2 \tau  \bm{b}_n \times (\bm{P}(t)\times \bm{b}_n) +\bm{G} + D_S \nabla^2 \bm{P} + \frac{\bm{J}}{ne}\cdot \nabla \bm{P},
\end{eqnarray}
\end{widetext}
which is identical to Eq. (\ref{eq:main}) except for the addition of the last two terms which describe spin diffusion and spin drift.

The physical device geometry was cast into a 1D finite-element model, where spin may drift and diffuse
laterally in the sample plane. The simplification to 1D is appropriate at cryogenic temperatures given
the device aspect ratio, where the spin diffusion length in GaAs is larger than the channel thickness.
Eq. \ref{eq:spinDiffusion} is iterated forward until steady state ($\frac{d \bm{P}}{dt}=0$) is
reached. The standard form for the Overhauser field\cite{Chan2009} is used to calculate $\bm{b}_n$ at
each spatial coordinate. Upon solving for the steady-state spatially dependent spin polarization in
the channel at each applied field, the 3T Hanle signal $\Delta V_{3T}$ is extracted by projecting the
spin polarization at the injector contact $\bm{P}_{inj}$ onto the magnetization of the injector
ferromagnet $\bm{M}$, \cite{Lou2006} i.e $\Delta V_{3T} \propto \bm{P_{inj}} \cdot \bm{M}$. A single
overall scaling factor is applied to compare the model to the data.

In Figure \ref{fig:angleExp}, the results of measuring the oblique angle dependence of the 3T Hanle
signal at 60 K are shown, along with the corresponding fits to the model described above. For
comparison, the effects of adding the anisotropic hyperfine relaxation terms discussed previously are
shown side-by-side with the fits without the anisotropic hyperfine relaxation terms.  In both cases, a
single set of parameters are used to fit the data at all angles. The results show that adding the
anisotropic hyperfine terms noticeably improve the fitting of the Overhauser peak at large oblique
angles, for which diffusion alone systematically overestimates the magnitude and underestimates the
width of the satellite peak. Without anisotropic relaxation, the height and width of the satellite are
determined only by the spatial variation of the Overhauser field on the scale of the electron spin
diffusion length.  This mechanism alone, however, is not sufficient to explain the broadening and
suppression at larger oblique angles. Inclusion of the additional smaller length-scale nuclear field inhomogeneity, via the anisotropic term, further reduces and broadens the Overhauser peaks. Table I contains the parameters used to fit the 3T signal both
without and with the anisotropic hyperfine terms.  Note that the addition of the anisotropic mechanism
does not change the other fitting parameters significantly, and the isotropic lifetime is essentially
unchanged. Measurements were also taken as a function of  injection bias current at fixed angle.  The fits to the model in this case are comparable to those for the angle dependence at fixed bias. Discrepancy between model and experiment is attributed to a $\pm$ 1$^{\circ}$ uncertainty in the angle of field with respect to sample. Additionally the obtained small values for $k_n$ and $k_0$ are only on the edge of the strong motional narrowing approximation.



\begin{table}[h]
\begin{tabular}{ | l || c | r |}
 \hline
  Parameter & w/o aniso. term & w/ aniso. term \\ \hline
  $\tau_s$ & $3.3$ ns & $3.4$ ns \\ \hline
  $b_{nuc}$ & $-1.50\times 10^4$ Oe & $-1.67\times 10^4$ Oe \\  \hline
  $b_e$ & $-82$ Oe & $-73$ Oe \\ \hline
  $\sqrt{\xi}B_L$ & $104$ Oe & $95$ Oe \\ \hline
  $k_0$ & $--$ & $2.1$ ns$^{-1}$ \\ \hline
  $k_n$ & $--$ & $0.45$ ns$^{-1}$ \\  \hline
\end{tabular}
\caption[]{Fitting parameters for curves in Figure \ref{fig:angleExp}.}
\end{table}

\begin{figure}[ptbh]
    \begin{centering}
        \includegraphics[scale = 1,trim = 0 0 0 0, angle = -0,clip]{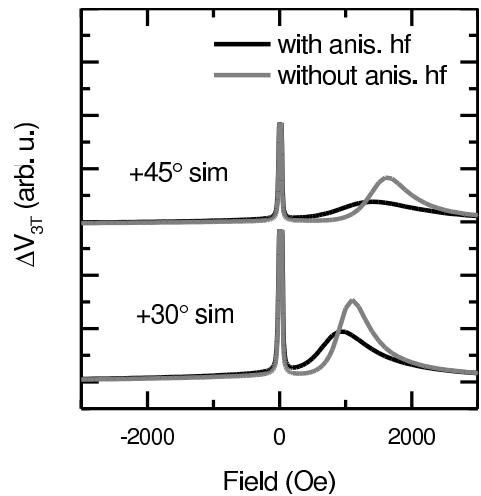}
        \caption[]{Large angle solutions to the spin diffusion model, Eq. (\ref{eq:spinDiffusion}), with and without the anisotropic spin relaxation terms.}\label{fig:diffusion}
    \end{centering}
\end{figure}

Oblique angles larger than $\pm 20 \degree$
were experimentally inaccessible due to the switching of the ferromagnetic contact when the in-plane
component of the field reached the coercive field.
Figure \ref{fig:diffusion} shows the solutions of Eq. (\ref{eq:spinDiffusion}) for two larger angles, 30$^{\circ}$ and 45$^{\circ}$.
The trend followed at these higher angles is similar to what is viewed at the lower ones -- the anisotropic terms tend to decrease the magnitude of the Overhauser peak (black) when compared to their exclusion (red).  If ferromagnetic contacts with larger coercivity are available, a more rigorous test of the predictions of this theory will be possible.

Now we consider how the anisotropic mechanism may also be evident in optical spin injection experiments.\cite{Meier1984, Kikkawa2000, Ou2015} In these experiments, the nuclear field is extracted by taking the difference of the total precession frequency and the precession frequency due solely to the applied field.\cite{Kawakami2001}
As we have discussed here, due to the inherent inhomogeneity of the nuclear field, the inferred nuclear field is actually an \emph{average} nuclear field in the probed macroscopic optical spot size. From Eq. (\ref{eq:main}), the inferred nuclear field is then $\overline{ \bm{B}_n} = k_0 \bm{b}_n/(k_0 + k_n)$ which leads to the anisotropic term being
\begin{equation}
-\gamma^2 \frac{(k_0 +k_n)^2}{k_0^2}\tau  \overline{ \bm{B}_n}  \times (\bm{P}(t)\times \overline{ \bm{B}_n} ).
\end{equation}
We predict this term to be observable in time-resolved Faraday or Kerr rotation experiments.

\emph{Conclusions}. --- The influence of DNP on spin evolution in semiconductors has been observed for many years. However the inherent inhomogeneity of the large nuclear fields has been neglected as a spin relaxation process. We have shown that the nuclear field inhomogeneity leads to an anisotropic spin relaxation mechanism and we have demonstrated that this new mechanism can account for the oblique Hanle measurements for electrical spin injection into $n$-GaAs.

\emph{Acknowledgements}. --- This work was supported in part by C-SPIN, one of six centers of STARnet, a Semiconductor Research Corporation program, sponsored by MARCO and DARPA, and by NSF under DMR-1104951.

\end{document}